\documentclass[final,1p,11pt]{elsarticle}

\usepackage{epsfig}
\usepackage{array,tabularx,epsfig,mathrsfs,graphicx,rotating}
\usepackage{ifthen}
\usepackage{amsfonts}
\usepackage{ragged2e}
\PassOptionsToPackage{hyphens}{url}
\usepackage[hyphens]{url}
\usepackage{hyperref}
\usepackage{listings}
\usepackage{lineno}
\usepackage{subfig}
\usepackage{epstopdf}
\usepackage{color}
\usepackage{float}
\usepackage{verbatim}
\usepackage{color,soul}

\usepackage[normalem]{ulem} 
\usepackage{soul} 
\usepackage{amsmath,amssymb}

\let\originallesssim\lesssim
\let\originalgtrsim\gtrsim

\DeclareRobustCommand{\lesssim}{%
  \mathrel{\mathpalette\lowersim\originallesssim}%
}
\DeclareRobustCommand{\gtrsim}{%
  \mathrel{\mathpalette\lowersim\originalgtrsim}%
}

\makeatletter
\newcommand{\lowersim}[2]{%
  \sbox\z@{$#1<$}%
  \raisebox{-\dimexpr\height-\ht\z@}{$\m@th#1#2$}%
}
\makeatother

\hypersetup{
  colorlinks=true,
  linkcolor=blue,
  citecolor=blue,
  urlcolor=blue
}

\graphicspath{{figs/}}

\newcommand{\beq}{\begin{equation}}
\newcommand{\eeq}{\end{equation}}

\newcommand{\mjj}{m_{jj}}
\newcommand{\mbb}{m_{bb}}

\chardef\til=126

\journal{ANL-HEP-17239}
\date{Nov 20, 2021}

\begin{document}
\definecolor{mygreen}{rgb}{0,0.6,0} \definecolor{mygray}{rgb}{0.5,0.5,0.5} \definecolor{mymauve}{rgb}{0.58,0,0.82}

\lstset{ %
 backgroundcolor=\color{white},   
 basicstyle=\footnotesize,        
 breakatwhitespace=false,         
 breaklines=true,                 
 captionpos=b,                    
 commentstyle=\color{mygreen},    
 deletekeywords={...},            
 escapeinside={\%*}{*)},          
 extendedchars=true,              
 keepspaces=true,                 
 frame=tb,
 keywordstyle=\color{blue},       
 language=Python,                 
 otherkeywords={*,...},            
 rulecolor=\color{black},         
 showspaces=false,                
 showstringspaces=false,          
 showtabs=false,                  
 stepnumber=2,                    
 stringstyle=\color{mymauve},     
 tabsize=2,                        
 title=\lstname,                   
 numberstyle=\footnotesize,
 basicstyle=\small,
 basewidth={0.5em,0.5em}
}


\begin{frontmatter}

\title{Event-based anomaly detection for new physics searches at the LHC 
using machine learning}

\author[add1]{S.V.~Chekanov}
\ead{chekanov@anl.gov}

\author[add1]{W.Hopkins}
\ead{whopkins@anl.gov}

\address[add1]{
HEP Division, Argonne National Laboratory,
9700 S.~Cass Avenue,
Lemont, IL 60439, USA.
}

\begin{abstract}
This paper discusses model-agnostic searches for new physics at the Large Hadron Collider (LHC) using anomaly-detection 
techniques for the identification of event signatures that deviate from the Standard Model (SM). 
We investigate anomaly detection in the context of 
machine-learning approaches using autoencoders, and illustrate expected shapes of invariant masses in the outlier region using Monte Carlo simulations. 
Challenges and conceptual limitations of this approach are discussed.
\end{abstract}

\end{frontmatter}

\section{Introduction}
Searches for new physics at the Large Hadron Collider (LHC) are typically 
performed assuming model-dependent techniques 
based on (1) generations of Monte Carlo (``background'') events according to Standard Model (SM) predictions
and (2) phenomenological models beyond the Standard Model (BSM) for expected ``signal events''. 
Such simulations help guide analyzers to define the kinematic regions that are affected by BSM physics.
Observations of statistical deviations from the predicted SM background events
can signify the presence of the new physics predicted by such models.

The classic examples of the usage of the techniques mentioned above are searches for the  Higgs boson \cite{Aad:2012tfa,Chatrchyan_2012}. However,  the success of this approach is more difficult to quantify for BSM  searches since no new physics has been discovered in the last decades. 
It is not unreasonable to think that this technique may slow down the pace of discoveries compared to the previous 
decades when Monte Carlo (MC) event generators were not widely used. For example,
the model-specific searches guided by BSM simulations 
may  ``lock'' the attention of analyzers to a limited parameter domain of some narrowly-designed  models, while BSM physics may appear in regions that have never been anticipated, or have a subtlety in its signatures that cannot be easily uncovered by comparing data with SM Monte Carlo simulations 
(or are below the precision of such simulations).  

Anomaly detection in particle physics has recently been 
discussed in several papers \cite{var2019,2019ann,2019mass,pol2020anomaly,Aarrestad:2021oeb,fraser2021challenges,chekanov2021searches,jawahar2021improving,2021julia,hallin2021classifying, aguilarsaavedra2021anomaly, weakunsup, knapp2020adversarially}. 
As an example, the usage of variational autoencoders
trained on known SM processes that could help identify anomalous events  was discussed in the context of the inputs for selected classes of event signatures \cite{var2019}. This study explores a similar event-based anomaly detection method that utilizes machine learning (ML), but using the standard autoencoder with  the inputs defined without the assumptions about particular classes of SM or BSM events, i.e. are well suited for general searches.
This feature is important 
for model-independent searches for signal-like structures in invariant masses after applying the ML technique since such studies do not require complex simulations Monte Carlo events.

In this paper we show how a fraction of data (which may include BSM physics) can be used for training and identification of outlier regions. We also illustrate how the invariant mass of two jets ($\mjj$) in 
such outlier regions can be described.  Our usage of the ``anomalous`` region for searches in invariant masses is somewhat different compared to \cite{2019ann} where the neural networks are used
to detect data departures from a reference model. We will use a fit method to find possible deviations from the background, which is widely accepted method in many experimental publications. 
Finally, we will show that the shapes of the invariant masses are not expected to have signal-like distortions that can be misidentified as possible BSM signals.

\section{Strategy for BSM searches using anomaly detection}

We approach the question of  BSM searches using a model-agnostic method that builds upon 
a rather natural assumption that new physics may produce unexpected signatures (such as peaks in invariant masses) hidden in the large SM backgrounds. To find such BSM events, one can select uncharacteristic SM events (``outliers'') and look at their signatures. 
The used anomaly detection algorithm must not bias the signatures themselves  (i.e. should not create artificial peaks etc.). The proposed strategy is outlined below:

\begin{itemize}
\item 
Define an input (``feature'') space for ML, assuming that such a feature space is as general as possible, and can cover a large class of possible BSM signatures;
\item Apply an anomaly detection algorithm  to this feature space using statistical methods or ML;
\item Define anomalous events (``outliers''). This is the most ambiguous part of this technique that will be discussed later; 
\item Study of  physics distributions of the events in the outlier regions (``unblind'').
\end{itemize}
The last step can focus on distributions that may not require 
a precise knowledge of SM backgrounds. For example, one can simply look 
for evidence of contributions from resonant BSM phenomena. On a technical side, this requires the analysis of invariant masses. New states with two-body decays may introduce localized excesses in such distributions,
which can be found without using Monte Carlo simulations for background modeling.
In some cases when no model-independent features are expected, such as invariant-mass peaks,
one can use various control regions, or even simulations for SM events.

Input data for anomaly detection  algorithms should reflect the fact that collision events produce various particles  (or more complex objects, such as jets or $b$-jets).  The lists that hold the information about such particles have variable sizes, i.e. change from event to event. One possibility to deal with a varying-size list is to ``map'' it to the fixed-size data structures.  We will use  a method based on
the rapidity-mass matrix (RMM) \cite{Chekanov:2018nuh,Chekanov:2018zyv} designed to represent a large number of relatively uncorrelated single and two-particle densities as a fixed-size sparse matrix.
Due to the unambiguous mapping of many popular experimental signatures to the matrix values, the RMM transformation will allow the usage of a broad range of ML techniques.
In the case when no new physics is found in the outlier region, 
one significant advantage in using the variables included in RMM lies in the fact that the calculated 95\% confidence level limits on cross sections of possible new physics in this region 
will be approximately Lorentz-invariant with respect to boosts along the beam axis.  
Alternative to RMM inputs for ML, such as four-momentum in the detector frame, will be discussed as well.

\section{Monte Carlo simulations}

\subsection{Input for anomaly detection}

To  explore the anomaly detection method proposed in this paper, 
we used MC event simulations. They were taken from the previous studies \cite{Chekanov:2018zyv,chekanov2021modelindependent}. Below we give a brief summary of the definition of these samples.

We use the PYTHIA 8 MC model \cite{Sjostrand:2006za} for the generation of
$pp$-collision events at a center of mass energy of $\sqrt{s}=13$~TeV.
The NNPDF 2.3 LO \cite{Ball:2014uwa} parton density function
from the LHAPDF library \cite{Buckley:2014ana} was used.
Two SM processes were generated: (1) light-flavored QCD dijets, (2) vector and scalar boson production and $t\bar{t}$ production. A minimum value of 50~GeV on generated invariant masses of the $2\to2$ system was set. For each event category, all available sub-processes were simulated at leading-order QCD with parton showers and hadronization.  Stable particles
with a lifetime of more than $3\cdot 10^{-10}$ seconds were considered, while
neutrinos were excluded from consideration.
All decays of top quarks, $H$ and  vector bosons were allowed.

The charged Higgs boson process ($H^+t$)  using  the diagram  $bg\to H^+t$ for
models with two (or more) Higgs doublets \cite{Akeroyd:2016ymd} was used as a benchmark BSM model. All decays
of the top quark and $H^+$ were allowed.
In addition to this model, the sequential SM (SSM) and the simplified dark-matter model (DM) with $W$ production were used.
These BSM models were discussed in \cite{Chekanov:2018zyv}. The event simulations were created using the $Z'$ masses
in the range of 1 -- 6 TeV, generated with a step size of 0.5~TeV. 10k events were used for each $Z'$ mass.

Jets, isolated electrons and muons  were reconstructed from stable particles.
Jets were constructed with the anti-$k_T$ algorithm \cite{Cacciari:2008gp} as implemented in the {\sc FastJet} package~\cite{Cacciari:2011ma} with a distance parameter of $R=0.4$, which is typically used in the ATLAS experiment.
The minimum transverse energy of all jets was $40$~GeV in the pseudorapidity range of $|\eta|<2.5$. Leptons are required to be isolated using 
a cone of size $0.2$ in the azimuthal angle and pseudo-rapidity 
defined around the true direction of the lepton. All energies of particles inside this cone are summed. A lepton is considered to be isolated if it carries more than $90\%$ of the cone energy. The SM background processes require simulations of misidentification rates for muons and leptons (``fake rates``). We use a misidentification rate of 0.1\%  for muons, and 1\% for electrons \cite{Chekanov:2018zyv}.
This is implemented by assigning the probability of $10^{-3}$ ($10^{-2}$)
for a jet to be identified as a muon (electron) using a random number generator.
The distributions were obtained for events having at least one isolated
lepton with transverse momentum  $p_T^{l}>30$~GeV and two jets with  $p_T^{j}>30$~GeV.
The SM background MC samples are available from the HepSim repository \cite{Chekanov:2014fga}. 

The events were transformed to the RMMs with  five types ($T=5$) of the reconstructed objects:
jets ($j$), $b$-jets ($b$), muons ($\mu$), electrons ($e$) and photons ($\gamma$). Up to ten
particles per type  were considered ($N=10$), leading to the so-called T5N10 topology for the RMM inputs. 

In total, about 1~million Monte Carlo events for SM processes were generated. About 100k events were used for ML training.
This sample was divided into 70k events for training and 30k for validation.
The trained neural network was applied to 900k SM events and BSM samples to illustrate the separation
between these samples in the ``loss'' phase space of the trained neural network.

\subsection{ML architecture}

An autoencoder is used with the RMM inputs. The RMM input dimension is 51x51, thus leading to 2601 inputs. After removing cells with 0 cell values for all events, 750 input columns were retained. This number defines the number of input neurons. Note that the reduction of inputs  should be determined using the full
sample of events or even several BSM models, 
to avoid situations when events with large multiplicities do not fit in the input layer. 
Alternatively, a topology was tested with exactly 2601 inputs, keeping columns with 0 values for all events. In this case, the results were very similar since autoencoder dismisses inputs that have columns with 0 values. However, this method of training will require more computational memory to load the  ``wide'' input layer.

The scaling of input values to the range [0 -- 1] has been applied for comparisons with an alternative ML feature space to be considered later. For the RMM, such transformations of input data are unnecessary since the RMM values are already re-scaled by definition. Our tests indicate that the results are nearly identical when using the RMM inputs without
additional transformations. In addition, the standardization of the input was tested (before rescaling to the range [0 -- 1]), 
but the results were found to be similar.

The autoencoder is implemented in the {\tt Keras} package with the {\tt TensorFlow}  backend \cite{abadi2016tensorflow}. It compresses the inputs
into lower dimensions  and then decodes the data in order to 
reconstruct the original input. The latent (``bottleneck'')  layer with 5 neurons  holds the compressed representation of the input data.
The network had 2 hidden layers before
the latent layer, with 20 and 10 neurons (``encoder''). After the latent layer, 2 hidden layers with 20 and 10 neurons are added. The output layer has the same number of neurons as the input layer. This architecture leads to  about 32,000 trainable parameters.
The model used the efficient Adam optimizer to minimize the mean squared error between the input and decoded output.
The reconstruction errors are used as the anomaly scores. 

In addition, a network was tested by varying the number of neurons in the
hidden and latent layers. The outcome was found to be similar to the results of the ML architecture described above. 

The ReLU activation function was used for all layers. 
As a test, sigmoid activation was also used
for the input and output layers, but it was found that this choice leads to a slower training.
In addition, the performance (in terms of SM and BSM separation) is worse than for the ReLU activation.

\subsection{Training}

\begin{figure}
\begin{center}
\subfloat[Using SM events] {
   \includegraphics[width=0.45\textwidth]{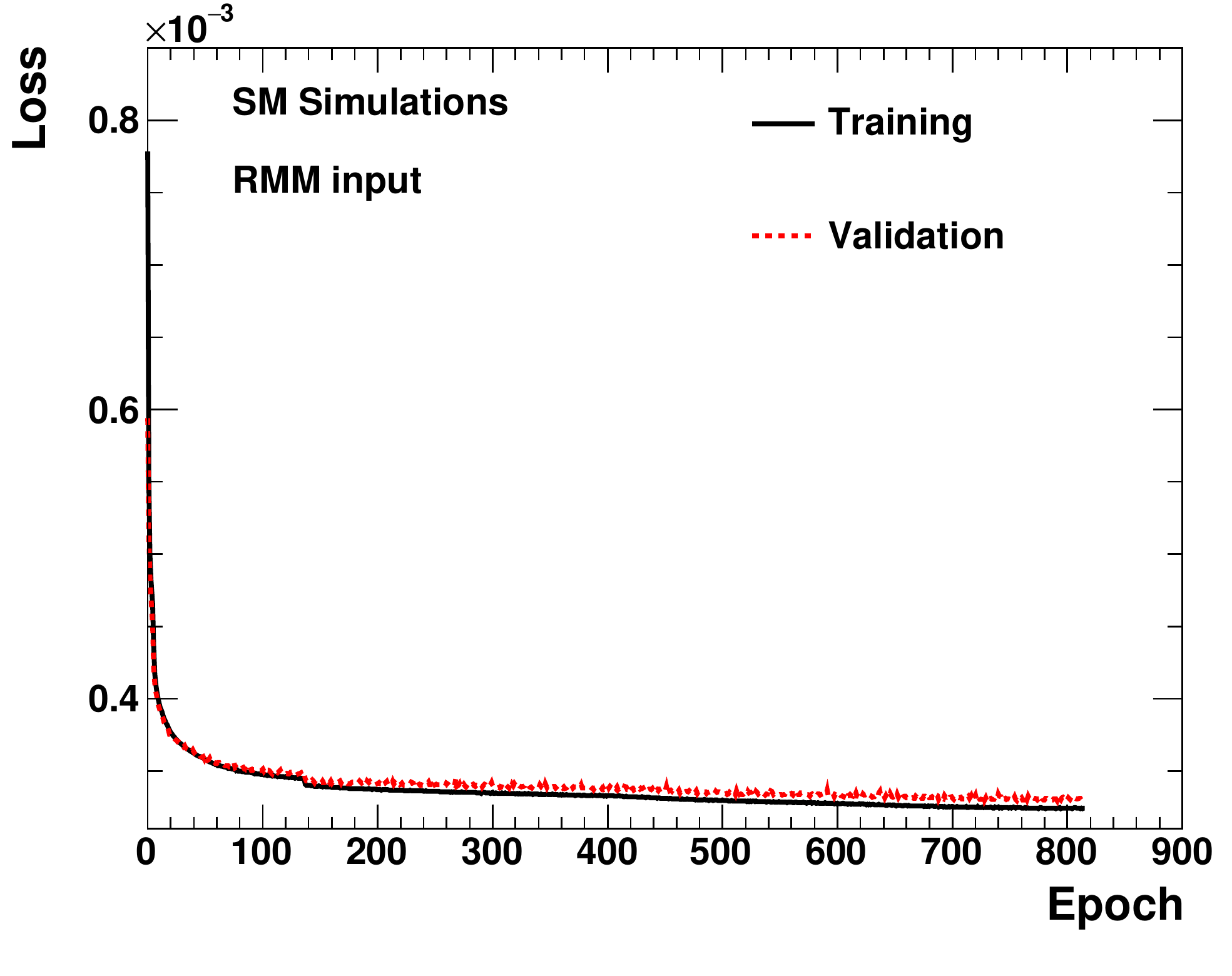}\hfill
   }
   \subfloat[Using SM plus BSM events] {
    \includegraphics[width=0.45\textwidth]{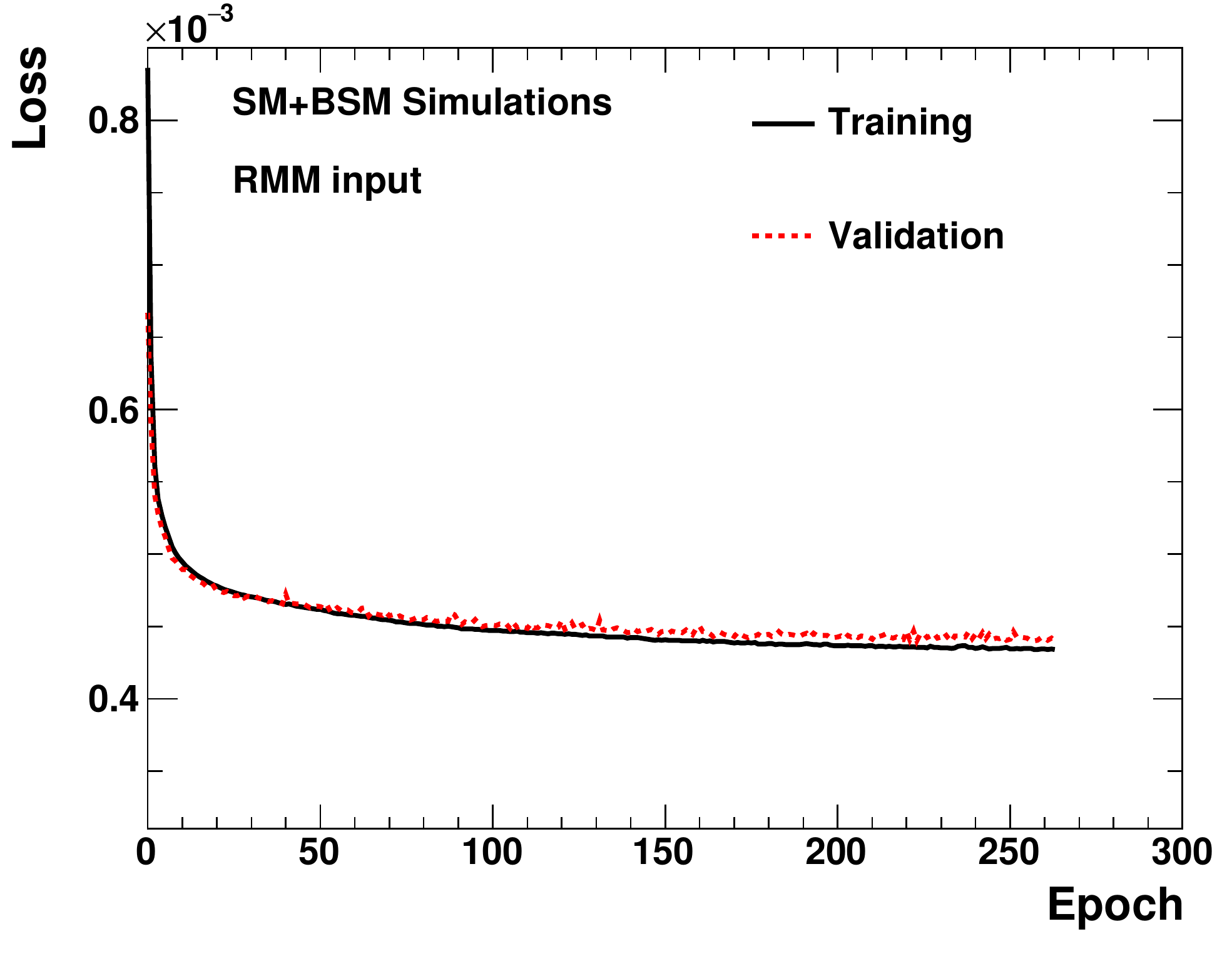}
    }
\end{center}
\caption{Loss as a function of epochs using the RMM inputs when using (a) only SM events; (b) using SM events plus 1,000 BSM events described in the text.
}
\label{fig:loss}
\end{figure}

The autoencoder training was performed on the SM Monte Carlo events that consist of multi-jet QCD events selected with at least one lepton, $t\bar{t}$ and $W$+jet events (added according to their rates). The selection cuts of MC objects 
were described above.  The training  batch size was 100.
The training is stopped when the validation sample stops showing improvements after  30 epochs.
Figure~\ref{fig:loss}(a) shows the values of the loss as a function of epochs.
This learning curve is used to diagnose the training, i.e., 
a model's learning performance over time. A good fit of the input data by the
autoencoder was observed.

In real situations, BSM models can contribute to a fraction of data used in training.
Figure~\ref{fig:loss}(b) shows the loss values
as a function of epochs for MC simulations using SM events plus additional BSM events (assuming 1,000 BSM events for each masses of $Z'/W'$ bosons in the range 1 -- 6 TeV, with a 1 TeV step size).
This number of events corresponds to a cross section of about 7~fb for the LHC Run2 data.
The results show that the loss values are larger compared to the SM-only training scenario. This is due to the fact that the SM+BSM events lead to a large number of non-zero RMM cells, compared to the SM-only events. The learning curve can be effectively minimized after a few hundred epochs, similar to the SM-only events used for the training.

\subsection{Results using ML}

\begin{figure}
\begin{center}
   \subfloat[Using SM events for training] {
   \includegraphics[width=0.45\textwidth]{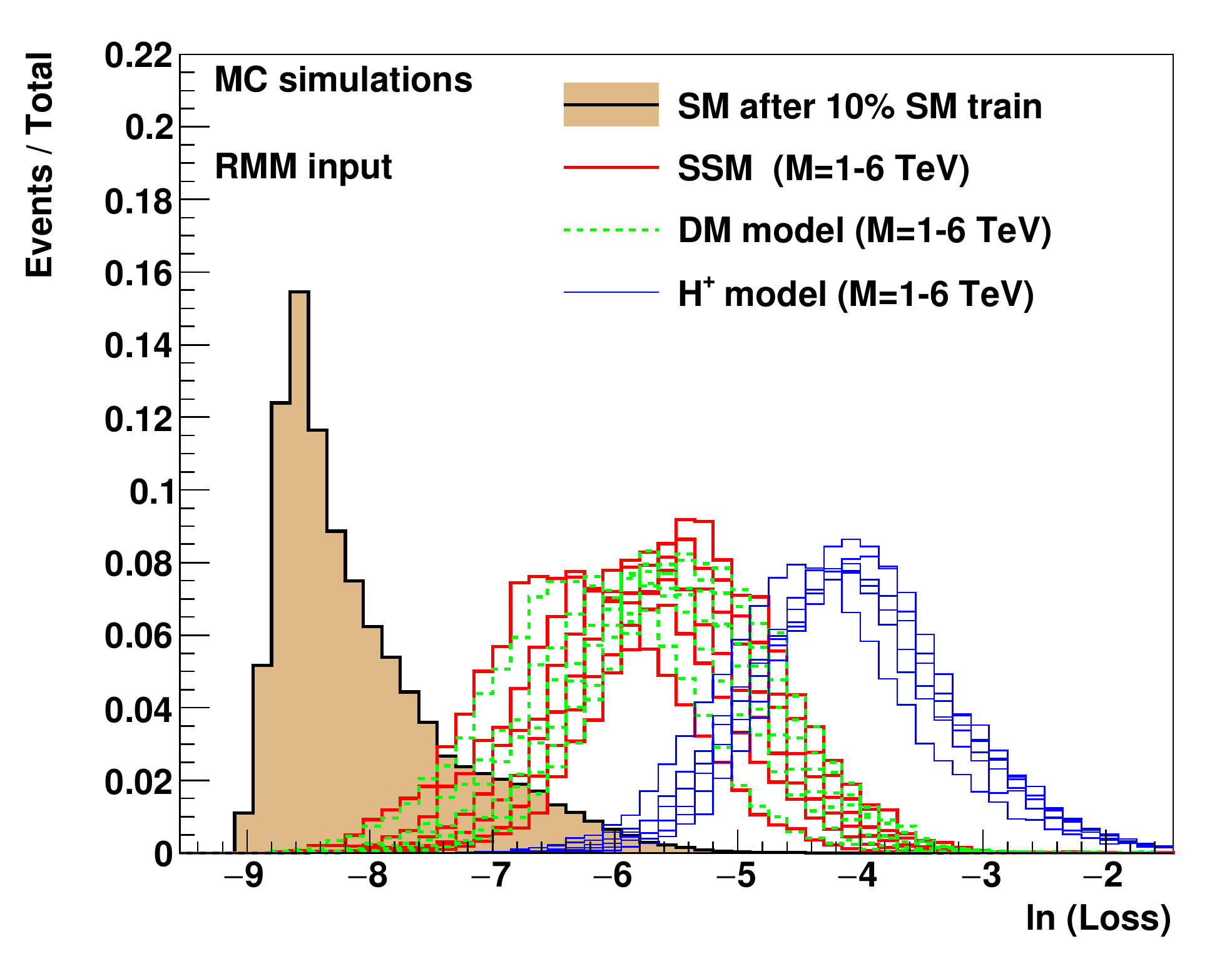}\hfill
   }
   \subfloat[Using SM+BSM for training] {
   \includegraphics[width=0.45\textwidth]{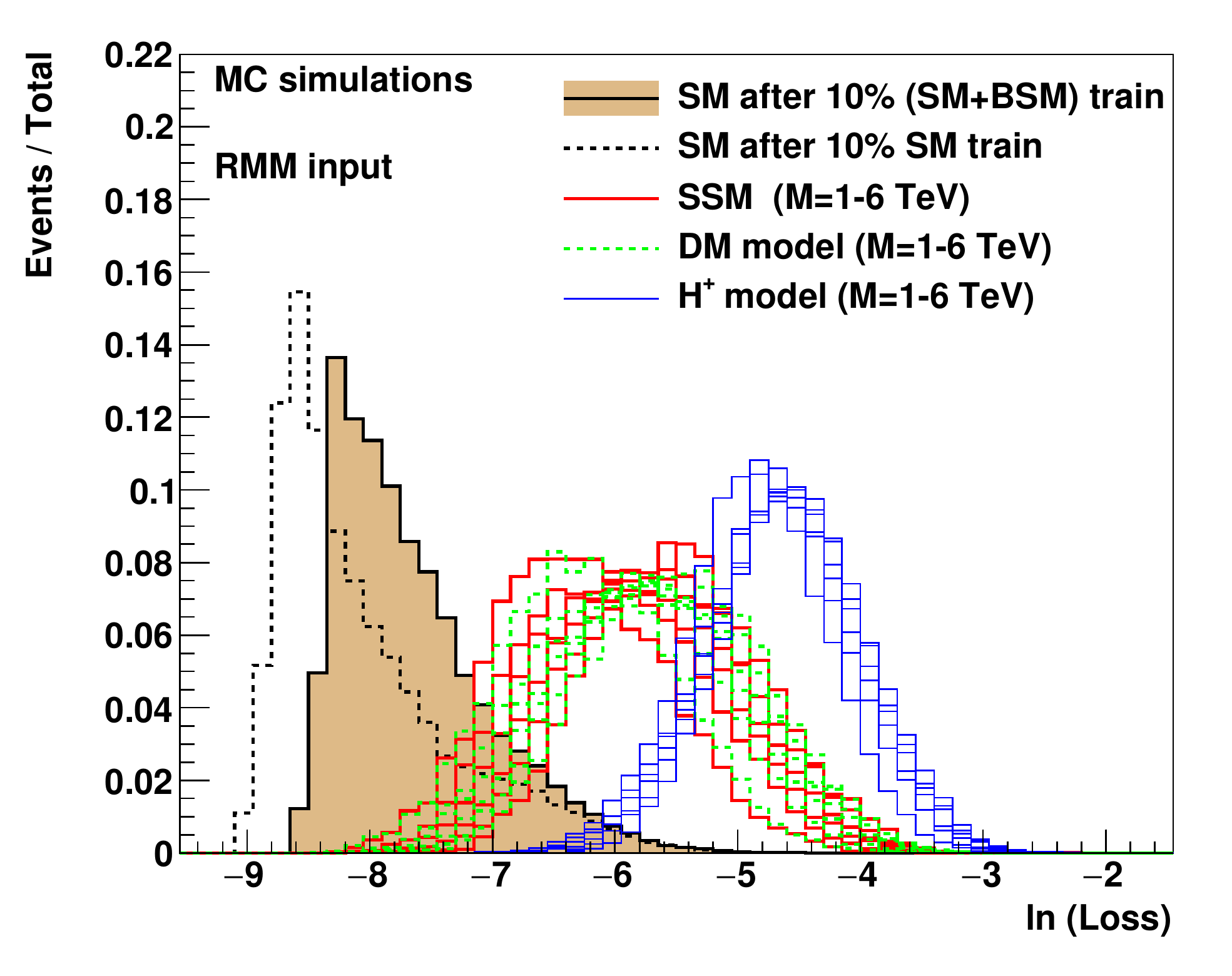}\hfill
   }
\end{center}
\caption{Distributions of the loss values for the trained autoencoder with  (a) 10\% of SM events and (b) 10\% of SM and BSM models.
The BSM models used in training had 1,000 events for each mass point in the range 1 -- 6 TeV. The BSM models are shown for $Z'/W'$ heavy bosons using the same mass  (shown with different lines of the same color). The distributions are normalized to 1.
The larger the mass of the resonance, the further away the line is from the SM distribution.
For the comparison, the black dashed line shows the SM events when using 10\% of the event samples without BSM events. }
\label{fig:resu}
\end{figure}

The trained autoencoder  was used to process the rest of the SM events (900k) and BSM events (that were not used in training).  The event samples were reconstructed with at least one lepton ($p_T^l>30$~GeV), before being
processed by the network.
Figure~\ref{fig:resu}(a) shows the loss values for these samples. It can be seen that there is a good separation between the SM events and the BSM events. 

As mentioned before,  BSM physics, if exists, may contribute to the data used in training. Therefore, it is important to verify the effect
of BSM events on the performance of the trained autoencoder. 
Figure~\ref{fig:resu}(b) shows the same MC simulations after applying the autoencoder trained using SM events plus additional BSM events (assuming 1,000 BSM events for each mass point used in the training).
For the comparison, the dashed black line shows the SM events when using 10\% of SM events for training (without BSM models). Some effect on the autoencoder from BSM models is observed, but the effect is not significant when it comes to the separation of the SM from BSM.
The largest effect from the inclusion of the BSM models in the training process was found on the BSM models themselves, rather than on the bulk of events dominated by the SM. 

One of the important questions concerning  anomaly detection is how to define the  
``anomalous'' region in terms of the $\ln (\mathrm{Loss})$ values.
One option is to define the upper value on BSM cross sections for a specific region
of the invariant mass of  two jets, $\mjj$, 
which will be analyzed in the outlier region. 
This cross section can be converted to
the number of collision events which can be used to define the selection cut 
$C$ on the $\ln (\mathrm{Loss})$ values. 
The cut $C$ will ensure that the outlier region will likely contain 
the BSM models without reducing their acceptance.
We do not define such a cut in this paper since it has to be estimated using BSM MC generators after a full detector simulation and acceptance cuts.
For the truth-level cross sections reported by the  BSM MC generators used in this paper, 
the selection  $\ln (\mathrm{Loss}) > C$ where $C$ is a value in the range $(-6, -5)$. 
Another option is to use published experimental limits obtained using  dijet masses, 
and calculate the upper range for such limits. This value can be 
transformed into the number of collision events that will define  the value $C$
for a given integrated luminosity. This ``data-driven''
selection of the outlier region makes sure that the outlier is sensitive to physics at the statistics precision level where data are not well explored.

It should be pointed out that a requirement on the loss value to separate SM events from BSM events does not directly translate to the equivalent signal-over-background separation for invariant masses.  This is because invariant-masses distributions for large masses of BSM particles (such as $Z'$) can be dominated by the SM background which has similar loss values as for these BSM models.
For example, SM events at $\ln (\mathrm{Loss}) > -6.5$ corresponds to low invariant masses of two jets, and such SM events can already be far away from the search mass region of $Z'$. Therefore, the signal-over-background ratios should  be evaluated on a case-by-case basis. 
Some of such cases will be shown later.

When using the proposed anomaly detection method it is important to  understand biases arising in invariant-mass distributions after applying a requirement on the reconstruction loss. In particular, one should answer the following question: can such a selection of outlier events 
create artificial structures that can be viewed by analyzers as possible evidence for signal events?

To answer this question, the next step is to verify that the outlier regions do not contain
any deviations in two-body invariant masses that can be interpreted as signals due to new BSM phenomena.
Therefore, the $\mjj$ were reconstructed in the outlier region with  $\ln(\mathrm{Loss})>-6.5$.
The following fit hypothesis \cite{Aaboud:2018fzt,Aad:2020kep} is used to establish the fact that the autoencoder selection does not introduce any spurious bumps: 
\begin{equation}
  f(x) = p_1 (1 - x)^{p_2} x^{p_3 + p_4\ln x + p_5 \ln^2 x},
\label{eq:function}
\end{equation}
where $x \equiv \mjj /\sqrt{s}$ and the $p_i$\ are five free parameters to be obtained from  likelihood or $\chi^2$ minimisation fits.

\begin{figure}
\begin{center}
   \subfloat[$\mjj$ in the outlier region] {
   \includegraphics[width=0.45\textwidth]{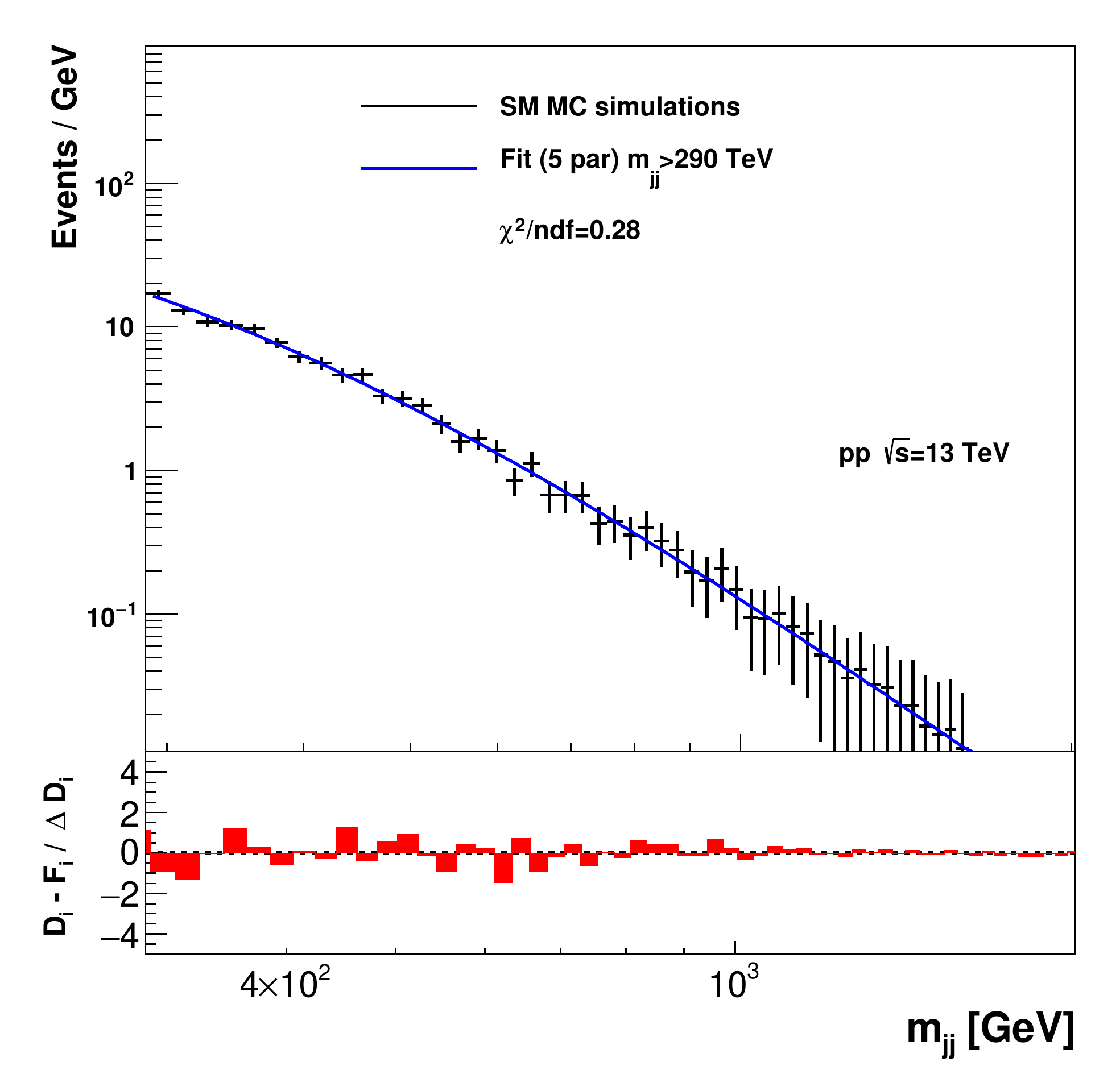}\hfill
   }
    \subfloat[$\mbb$ in the outlier region] {
      \includegraphics[width=0.45\textwidth]{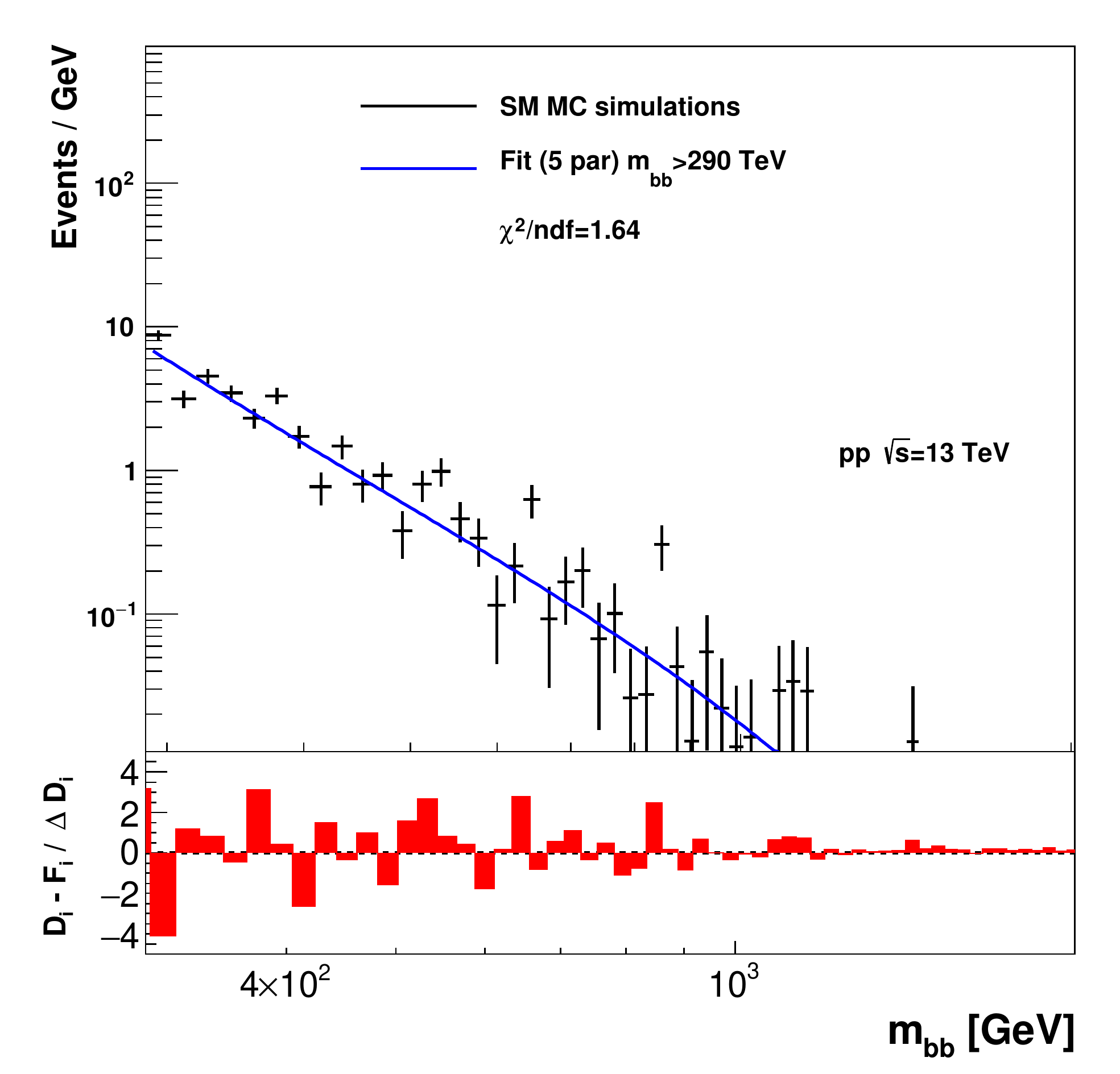}
   }
\end{center}
\caption{Invariant masses in the outlier region with  $\ln(\mathrm{Loss})>-6.5$.
The masses were reconstructed for light-favor jets and  $b$-jets using SM Monte Carlo simulations. The blue line is the fit function (see the text).
}
\label{fig:mass}
\end{figure}

Figure~\ref{fig:mass} shows the invariant masses of light-flavor jets and  $b$-jets 
in the SM MC simulations for the events
in the outlier region $\ln(\mathrm{Loss})>-6.5$ together with the $\chi^2/ndf$ fit using the analytic function shown above.
The bottom parts of the plots show the pull values of the fit. Good agreement with the function was found, without
significant deviations in the pulls. 
This study confirms that the outlier regions do not contain biases that may distort
the invariant masses (for the statistics of the MC events used in this paper).

\section{Using alternative inputs}
\label{inputs2}

We also tested other classes of the input variables for the ML algorithm, in addition to the RMM. As an example, we used the four-momentum variables $(P_x,P_y,P_z,E)$ for each object and missing transverse momentum. Such inputs, or similar inputs converted to the $(p_T, \eta, \phi, E$) space, were chosen in \cite{2019ann,jawahar2021improving,pol2020anomaly}. The approach \cite{jawahar2021improving,pol2020anomaly} of ``image-like'' fixed-size  data structures with  ordering in transverse momentum of jets/particles and zero-padding is almost identical to that proposed in \cite{Chekanov:2018nuh} for RMM.

In order to have a fixed size input for $(P_x,P_y,P_z,E)$ inputs, we considered up to 10 jets/particles per event (i.e., similar to the RMM) plus the missing momentum. In the case where an event does not fill all 10 positions, we fill the missing values with 0. This leads to the arrays with 
$5\times (10\times 4)+1=201$ values. 
As in the case of RMM, many values of these arrays have 0 values for all events.
Then the input was re-scaled to the range [0 -- 1]. The autoencoder had 20, 10 neurons for the hidden layers and 5 neurons for the latent layers. The input and output layers had 201 neurons. As before, ReLU activation was used.
The training is performed using 10\% of events with the SM and BSM models. The training is terminated when the validation sample stops showing improvements after 30 epochs.

\begin{figure}
\begin{center}
   \subfloat[Using SM+BSM for training] {
    \includegraphics[width=0.45\textwidth]{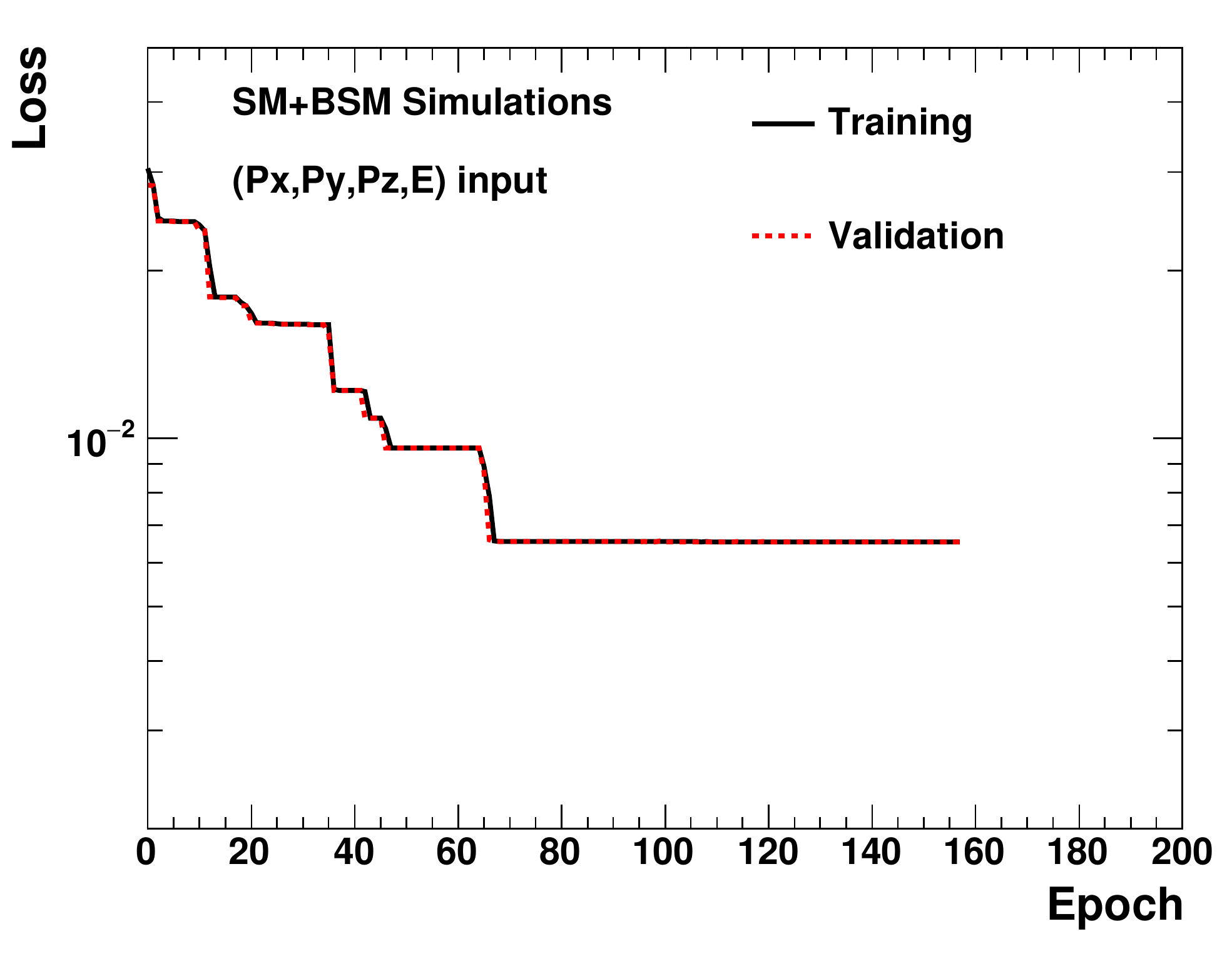}\hfill
   }
    \subfloat[Result of the trained network] {
     \includegraphics[width=0.45\textwidth]{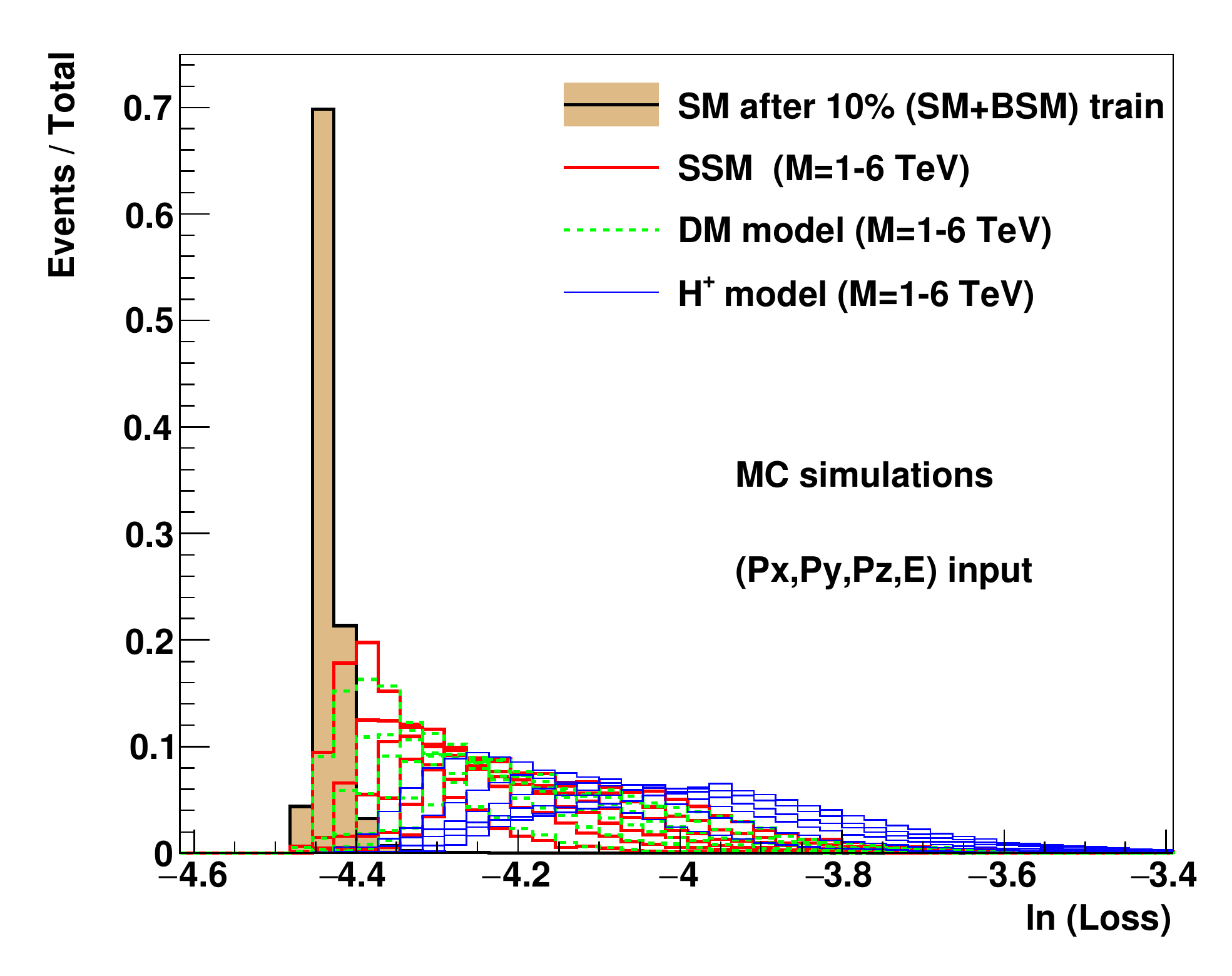}
   }
   
\end{center}
\caption{The loss values as a function of epochs using arrays with four-momenta using training of the autoencoder. The BSM models used for training had 1,000 events for each mass point in the range 1 -- 6 TeV. The right-hand figure shows the results of the trained network applied to the SM and BSM models. The BSM models are shown for $Z'/W'$ heavy bosons in the mass range 1 -- 6 TeV (shown with different lines of the same color).
}
\label{fig:alt}
\end{figure}

Figure~\ref{fig:alt}(a) shows the loss values as a function of epochs during training. It can be seen that improvements in the loss values have steps, unlike a smooth decrease in the case of the RMM.
This may indicate the lack of stability in the training. Repeating the training showed 
that the exact epoch at which the training should be stopped is hard to  reproduce.
The results of the training are shown in Figure~\ref{fig:alt}(b). This figure should be compared with the Fig.~\ref{fig:resu}(b) in the case of the RMM. One can see that the separation between SM and BSM models is possible when using a list of jets/particles with four-momentum, but the latter shows a narrower numeric range for the loss values. Since the SM shows the very narrow loss distribution, close
to the BSM signal models, this can be a disadvantage since the exact overlap of SM and BSM is more difficult to predict assuming realistic uncertainties for this distribution.
For example, a repeated training showed the height of the narrow SM histogram randomly changes in the range 0.4 -- 0.7.

It is worth mentioning that the compression factor for the four-momentum input is more than a factor four smaller than for the RMM inputs since the same ML architecture was used for a lesser number of the inputs (201 versus 750 in the case of the RMM after the 0-value trimming).
Reducing the number of neurons in the layer by a factor 2 leads to a sharper peak for the SM,
and more unstable results.
Although it is conceivable  to obtain good training results for the four-inputs, as shown in Fig.~\ref{fig:resu}(b), which is likely due to variations in the multiplicity of objects affecting the number of columns with zero-padding,
the lack of a stable behavior and reproducibility are the main disadvantages of  the four-momentum inputs. 
A possible reason for this training instability is in the difficulties to predict 
the mean of the Lorentzian-shaped distributions of the $P_x$, $P_y$ and $P_z$ variables. 
The autoencoder learns the peaks at 0.5 (after the re-scaling to the [0 -- 1] range),
but not the precise shape of the fast falling ``wings'' of such distributions, which are affected by the trivial symmetry in the azimuthal angle that introduces additional smearing of relevant features.
In contrast, the  RMM values have smoothly falling distributions which can be easier reproduced by the autoencoder. 
Other advantages of the RMM are automatic rescaling and normalization, visual debugging of inputs in the form of images and an approximate Lorentz
invariance with respect to the boosts along the beam axis. 

As mentioned earlier, another possible feature space is the fixed-size data structures with  $(p_T,\eta,\phi,E$) variables for each object after applying
the zero-padding, or some variation of  ``pick and choose'' variables. Three variables, $E$, $p_T$ and  $\eta$  belong to the RMM data structure. The $\phi$ variable is not too useful due to the flat distribution of the azimuthal angles in the detector frame\footnote{Indirectly,  the RMM includes the information about the difference $\phi_1-\phi_2$ via the two-body invariant masses.}.

We did not consider particular sets of ``pick and chose'' variables discussed in the literature cited before since such selections suffer from the lack of generality, and are usually considered for particular types of BSM signatures.

\section{Example for wide resonances}
\label{test1}

As a further example, we consider the situation where BSM models do not show sharp peaks in invariant masses, 
therefore, the outlier region should be studied using Monte Carlo simulations. 
For example, the $H^+$ model does not exhibit sharp peaks in invariant masses, thus it is a special case compared to the previous discussion.  We assumed that the mass of the $H^+$ boson is 0.5~TeV. 10,000 $H^+$ events were  generated as described above in all  decay modes. As before, one million SM events were created using PYTHIA 8 (selected with an isolated lepton) were used. 

\begin{figure}
\begin{center}
   \subfloat[Using SM events for training] {
   \includegraphics[width=0.45\textwidth]{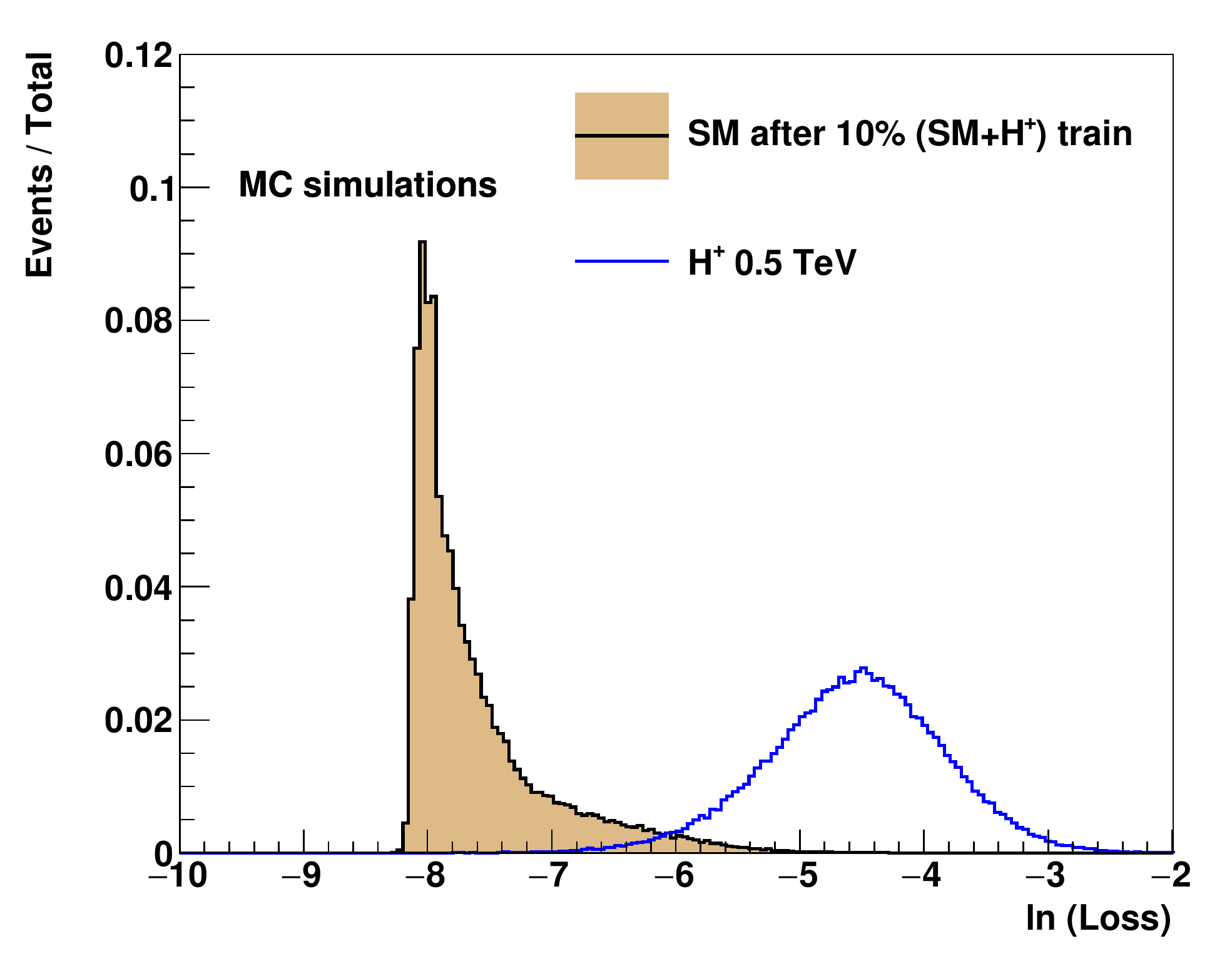}\hfill
   }
\end{center}
\caption{Distribution of the loss values for SM and $H^+$ events using 10\% of SM events and 10\% of 
$H^+$ for training (see the text).}
\label{fig:resuHP}
\end{figure}

We prepared a sample of simulated 
data using 10\% of the available events  (i.e. 1k for $H^+$ and 100k randomly selected SM events), converted to RMMs, 
and trained the same autoencoder as described previously. Figure~\ref{fig:resuHP} shows the loss values for the trained network.
A good separation between the $H^+$ events  and the SM simulation is observed.

\begin{figure}
\begin{center}
   \includegraphics[width=0.45\textwidth]{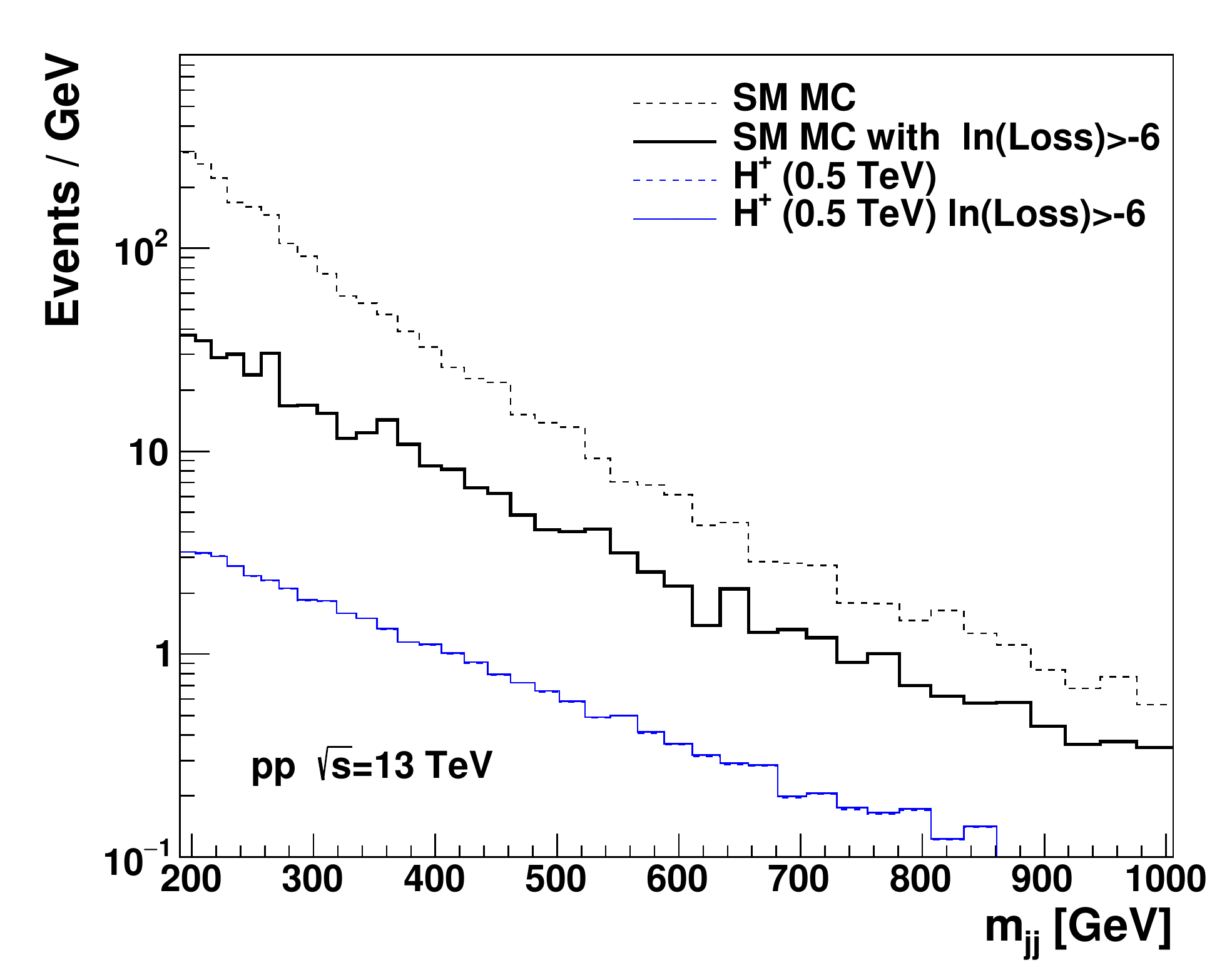}\hfill
   \includegraphics[width=0.45\textwidth]{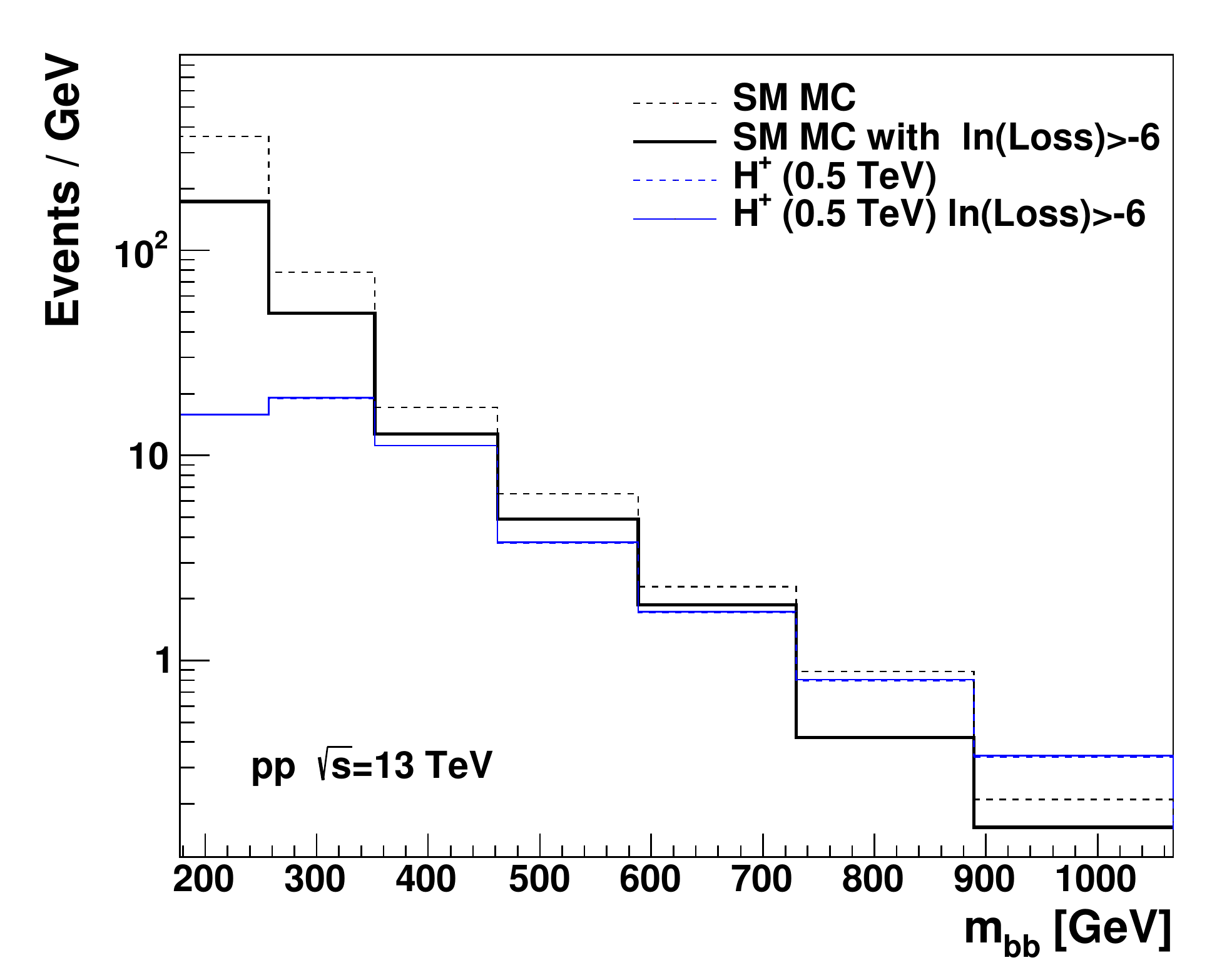}\hfill
\end{center}
\caption{The distributions of the invariant masses in the outlier region with  $\ln(\mathrm{Loss})>-6$.
The masses were reconstructed for light-favor jets (left) and for $b$-jets (right) using SM Monte Carlo simulations. The blue line is the fit function (see the text). 
}
\label{fig:massH}
\end{figure}

The invariant masses for SM and $H^+$ were reconstructed before and
after the requirement $\ln(\mathrm{Loss})>-6$. 
Figure~\ref{fig:massH} shows that the selection cut can significantly increase the signal-over background ratio. For large invariant masses of two $b$-jets, $\mbb$, the $H^+$ events dominate the distribution.

\section{Discussion}

The most important question for the anomaly-detection style analysis follows: what does it mean to observe new physics in the outlier events, assuming that the reliance on SM backgrounds and signal Monte Carlo simulations is not strictly required? 
One of the most straightforward ways to detect a signal without knowing the shape and normalization of the SM background is to observe signal-like enhancements on smoothly falling background distributions. 
One can construct 14 invariant masses from the five objects (as used in this analysis)
by selecting  objects with the largest transverse energies using the RMM inputs.
This ``bump-hunter'' approach does not rely on the precise knowledge of SM backgrounds, which is expected to be described by a smoothly falling curve. On a technical side, this would require an analysis of the top-right corner of the RMMs in the outlier sample.

It should be noted that, since a significant number of invariant masses should be inspected, the criteria for an observation of new physics 
should be higher than for the standard observations of  ``bumps'' in histograms. The number of invariant masses that are included
in the right-top corner of the RMM can be as large as several hundred. Therefore, the look-elsewhere-effect contributing to the statistical significance of an observation can be non-negligible.
The strength of this effect should be calculated depending on how many invariant masses will be analyzed.
The most sensible approach is to start with the analysis of several invariant masses on the left side of the RMM columns (for a given object type), which corresponds to objects with the largest transverse momentum. 

One of the main conclusions of this paper is that applying a selection based on the trained autoencoder with the RMM inputs does not bias  the shapes of invariant masses in the outlier region. This feature can be used
for model-independent searches of signals in invariant masses of the outlier region. Therefore, the studies presented in this paper make a strong case for using RMM with autoencoders for general event-based anomaly detection in invariant masses.

As we discussed in Sect.~\ref{test1}, enhancements in invariant masses may not be always reliable  signatures to detect new physics since BSM particles may have too broad widths. In this case,
SM Monte Carlo models can be utilized to compare with the rate of events in the outlier region, or even a simple visual event scan of the outlier region can be adopted.

\section*{Acknowledgments}
We gratefully acknowledge the computing resources provided by
the Laboratory Computing Resource Center at Argonne National Laboratory.
The submitted manuscript has been created by UChicago Argonne, LLC, Operator of Argonne National Laboratory (“Argonne”). Argonne, a U.S. 
Department of Energy Office of Science laboratory, is operated under Contract No. DE-AC02-06CH11357. The U.S. Government retains for itself, 
and others acting on its behalf, a paid-up nonexclusive, irrevocable worldwide license in said article to reproduce, prepare derivative works, 
distribute copies to the public, and perform publicly and display publicly, by or on behalf of the Government.
The Department of Energy will provide public access to these results of federally sponsored research in accordance with the 
DOE Public Access Plan. \url{http://energy.gov/downloads/doe-public-access-plan}. Argonne National Laboratory’s work was 
funded by the U.S. Department of Energy, Office of High Energy Physics under contract DE-AC02-06CH11357.

\section*{References}
\bibliographystyle{elsarticle-num}
\bibliography{references}

\end{document}